# The Baikal-GVD neutrino telescope: search for high-energy cascades

V.A. Allakhverdyan[1], A.D. Avrorin[2], A.V. Avrorin[2], V.M. Aynutdinov[2], R. Bannasch[3], Z. Bardačová[4], I.A. Belolaptikov[1], I.V. Borina[1], V.B. Brudanin[1], N.M. Budnev[5], V.Y. Dik[1], G.V. Domogatsky[2], A.A. Doroshenko[2], R. Dvornický[1,4], A.N. Dyachok[5], Zh.-A.M. Dzhilkibaev[2,*], E. Eckerová[4], T.V. Elzhov[1], L. Fajt[6], S.V. Fialkovsky[7], A.R. Gafarov[5], K.V. Golubkov[2], N.S. Gorshkov[1], T.I. Gress[5], M.S. Katulin[1], K.G. Kebkal[3], O.G. Kebkal[3], E.V. Khramov[1], M.M. Kolbin[1], K.V. Konischev[1], K.A. Kopański[8], A.V. Korobchenko[1], A.P. Koshechkin[2], V.A. Kozhin[9], M.V. Kruglov[1], M.K. Kryukov[2], V.F. Kulepov[7], Pa. Malecki[8], Y.M. Malyshkin[1], M.B. Milenin[2], R.R. Mirgazov[5], D.V. Naumov[1], V. Nazari[1], W. Noga[8], D.P. Petukhov[2], E.N. Pliskovsky[1], M.I. Rozanov[10], V.D. Rushay[1], E.V. Ryabov[5], G.B. Safronov[2], B.A. Shaybonov[1], M.D. Shelepov[2], F.Šimkovic[1,4,6], A.E. Sirenko[1], A.V. Skurikhin[9], A.G. Solovjev[1], M.N. Sorokovikov[1], I. Štekl[6], A.P. Stromakov[2], E.O. Sushenok[1], O.V. Suvorova[2], V.A. Tabolenko[5], B.A. Tarashansky[5], Y.V. Yablokova[1], S.A. Yakovlev[3], and D.N. Zaborov[2], Yu.A. Kovalev[11], Yu.Yu. Kovalev[11,16,17], A.V. Plavin[11,16], S.V. Troitsky[2], A.K. Erkenov[12], T.V. Mufakharov[13,18], Yu.V. Sotnikova[12], T. Hovatta[14,19], S. Kiehlmann[15,20] and A.C.S. Readhead[16]

[1] *Joint Institute for Nuclear Research, Dubna, Russia, 141980;* [2] *Institute for Nuclear Research, Russian Academy of Sciences, Moscow, Russia, 117312;* [3] *EvoLogics GmbH, Berlin, Germany, 13355;* [4] *Comenius University, Bratislava, Slovakia, 81499;* [5] *Irkutsk State University, Irkutsk, Russia, 664003;* [6] *Czech Technical University in Prague, Prague, Czech Republic, 16000;* [7] *Nizhny Novgorod State Technical University, Nizhny Novgorod, Russia, 603950;* [8] *Institute of Nuclear Physics of Polish Academy of Sciences (IFJ PAN), Krakow, Poland, 60179;* [9] *Skobeltsyn Institute of Nuclear Physics MSU, Moscow, Russia, 119991;* [10] *St. Petersburg State Marine Technical University, St. Petersburg, Russia, 190008;* [11] *Lebedev Physical Institute of the Russian Academy of Sciences, Moscow, Russia, 119991;* [12] *Special Astrophysical Observatory of RAS, Nizhny Arkhyz, Russia 369167;* [13] *Kazan Federal University, Kazan, Russia, 420008;* [14] *Finnish Centre for Astronomy with ESO, University of Turku, Turku, Finland, FI-20014;* [15] *Owens Valley Radio Observatory, California Institute of Technology, Pasadena, USA, CA 91125;* [16] *Moscow Institute of Physics and Technology, Dolgoprudny, Moscow region, Russia, 141700;* [17] *Max-Planck-Institut fur Radioastronomie, Bonn, Germany, 53121;* [18] *Shanghai Astronomical Observatory, Chinese Academy of Sciences, Shanghai, China, 200030;* [19] *Aalto University Metsahovi Radio Observatory, Kylmala, Finland, FI-02540;* [20] *Department of Physics, Univ. of Crete, Heraklion, Greece, GR-70013*

*E-mail: djilkib@yandex.ru

Baikal-GVD is a neutrino telescope currently under construction in Lake Baikal. GVD is formed by multi-megaton subarrays (clusters). The design of Baikal-GVD allows one to search for astrophysical neutrinos already at early phases of the array construction. We present here preliminary results of a search for high-energy neutrinos with GVD in 2019-2020.

________________________________

*Presente

1.     Introduction

The deep underwater neutrino telescope Baikal-GVD (Gigaton Volume Detector) is currently under construction in Lake Baikal [1]. Baikal-GVD is formed by a three-dimensional lattice of optical modules, which consist of photomultiplier tubes housed in transparent pressure spheres. They are arranged at vertical load-carrying cables to form strings. The telescope has a modular structure and consists of functionally independent clusters - sub-arrays comprising 8 strings. Each cluster is connected to the shore station by an individual electro-optical cable. The first cluster with reduced size, named "Dubna", was deployed and operated in 2015. In April 2016, this array has been upgraded to the baseline configuration of a GVD-cluster, which comprises 288 optical modules attached along 8 strings at depths from 750 m to 1275 m. In 2017-2021, seven additional GVD-clusters were commissioned, increasing the total number of optical modules up to 2304 OMs.

IceCube has discovered a diffuse flux of high-energy astrophysical neutrinos in 2013 [2]. The 7.5 year data sample of high-energy starting events (HESE) comprises 102 events, 75 of which are identified as cascades and 27 as track events [3]. The Baikal Collaboration has long-term experience with the NT200 array to search for diffuse neutrino fluxes via the cascade mode [4,5]. Baikal-GVD has the potential to record already at early phases of construction astrophysical neutrinos with a flux measured by IceCube [6]. One search strategy for high-energy neutrinos with Baikal-GVD is based on the selection of cascade events generated by neutrino interactions in the sensitive volume of the array [7]. Here we discuss the preliminary results based on data accumulated in 2019-2020.

2.     Cascade detection with GVD cluster

**2.1 Reconstruction method**

The procedure for reconstructing the parameters of high-energy showers - the shower energy, direction, and vertex - is performed in two steps. In the first step, the shower vertex coordinates are reconstructed using the time information from the telescope's triggered photo-sensors. In this case, the shower is assumed to be a point-like source of light. The $\chi^2$ minimization parameters are shower coordinates:

$$\chi_t^2 = \frac{1}{(N_{hit} - 4)} \sum_{i=1}^{N_{hit}} \frac{(T_i(\vec{r}_{sh}, t_0) - t_i)^2}{\sigma_{ti}^2},$$

where $t_i$ and $T_i$ are the measured and theoretically expected trigger times of the *i*th photo-sensor, $t_0$ - the shower generation time, $\sigma_{ti}$ - the uncertainty in measuring the time, and $N_{hit}$ is the hit multiplicity. The reconstruction quality can be increased by applying additional event selection criteria based on the limitation of the admissible values for the specially chosen parameters characterizing the events.

In the second step, shower energy and direction are reconstructed by applying the maximum-likelihood method and using the shower coordinates reconstructed in the first step. The values of the variables θ, φ, and $E_{sh}$ corresponding to the minimum value of the following functional are chosen as the polar and azimuth angles characterizing the direction and the shower energy:

$$L_A = -\sum_{i=1}^{N_{hit}} \ln p_i\left(A_i, E_{sh}, \vec{\Omega}_{sh}(\theta,\varphi)\right).$$

The functions $p_i(A_i, E_{sh}, \Omega_{sh}(\theta,\varphi))$ are the probabilities for a signal with amplitude $A_i$ (measured in photoelectrons) from a shower with energy $E_{sh}$ and direction $\Omega_{sh}$ to be recorded by the $i$th triggered photo-sensor:

$$p_i = \sum_{n=1}^{\infty} P(n/\bar{n}) \int_{A_i - \frac{\alpha}{2}}^{A_i + \frac{\alpha}{2}} \xi_i(A,n) dA,$$

where $P(n/\bar{n})$ is the probability of detecting $n$ photoelectrons at a mean $\bar{n}$ for the Poisson distribution, $\xi(A,n)$ is the probability density function for recording the amplitude $A$ at an exposure level of $n$ photoelectrons, and α is the scale-division value of the amplitude in photoelectrons. The mean $\bar{n}$ are determined by simulating the responses of the telescope's OMs to the Cherenkov radiation of a shower with energy $E_{sh}$ and direction $\Omega_{sh}$ with allowance made for the light propagation in water, the relative positions and orientation of the OMs and the shower, and the effective OM sensitivity.

## 2.2 Performance of GVD-cluster

The search for high-energy neutrinos with a GVD-cluster is based on the selection of cascade events generated by neutrino interactions in the sensitive volume of the array. Performances of event selection and cascade reconstruction procedures were tested by MC simulation of signal and background events and reconstruction parameters of cascades. After reconstruction of cascade vertex, energy and direction and applying quality cuts, events with a final multiplicity of hit OMs $N_{hit} \geq 20$ were selected as high-energy neutrino events. The requirement of high hit multiplicity allows substantial suppression of background events from atmospheric muon bundles. The accuracy of cascade energy reconstruction is about 10%-30% and the accuracy of direction reconstruction is about 2-4 degree (median value) depending on location and orientation of cascade. The vertex resolution is about 2 m [7]. Energy distributions of cascade events expected for one year observation from astrophysical flux following a power law $E^{-2.46}$ spectrum and single-flavour normalization $4.1 \times 10^{-6}$ GeV$^{-1}$ cm$^{-2}$ s$^{-1}$ sr$^{-1}$ [6, 8], as well as distribution of expected background shower events from atmospheric neutrinos are shown in Fig.1. The expected number of background events from atmospheric neutrinos is strongly suppressed for energies higher than 100 TeV. About 0.4-0.6 cascade events per year with energies above 100 TeV and hit multiplicities $N_{hit} > 20$ from astrophysical flux following a $E^{-2.46}$ spectrum and 0.08 background events from atmospheric neutrinos are expected.

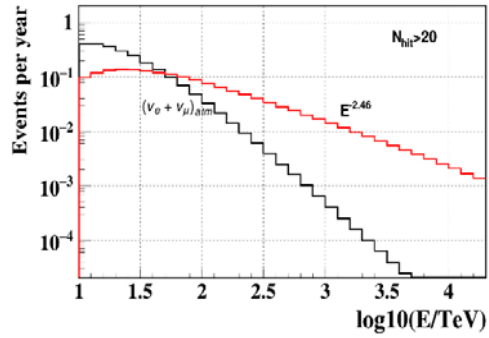

**Figure 1.** Energy distributions of events expected for one year observation from astrophysical flux with $E^{-2.46}$ spectrum and IceCube normalization (see text). Also shown is a distribution of expected background events from atmospheric neutrinos.

## 3. Data analysis and results

To search for high-energy neutrinos of astrophysical origin the data collected by five clusters in 2019 and by seven clusters in 2020 have been used. A data sample accumulated by the array trigger, corresponds to 2915 one cluster live days. After applying procedures of cascade vertex and energy reconstruction for hits with charge higher than 1.5 ph. el. and all selection cuts, 72 cascade-like events with OMs hit multiplicity $N_{hit} > 19$ and E > 40 TeV have been selected. The hit multiplicity dependence on reconstructed cascade energy is shown in Fig.2. Ten events were reconstructed as cascades with energies above 100 TeV and $N_{hit} > 19$. The energy and cosine of zenith angle distributions of selected events, as well as expected distributions of background events from atmospheric muons are shown in Fig.3, left and right panels, respectively. The reconstructed coordinates of selected events are shown in Fig.4.

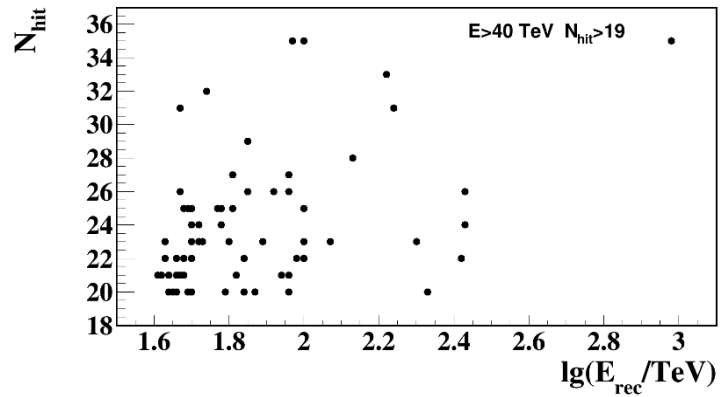

**Figure 2.** Hit multiplicity dependence on the reconstructed cascade energy.

The number of selected high-energy cascades, as well as their energy and zenith angular distributions are consistent with expectations from atmospheric muon flux. Detection of astrophysical diffuse neutrinos requires a further suppression of background events by factor of 2-3. Additional background suppression procedure is based on an analysis of OM hit time distributions in cascade-like events. The precision of OM hits resolution in GVD is about 20 ns or better. This allows to separate hits induced by reconstructed cascade and muons and identify background events from atmospheric muons. Three types of OM hits were defined: hits from cascade generated on muon track (type 1), hits induced by muon which generates cascade (type 2), hits induced by another muons of muon bundle (type 3). The cuts on number of type 2 and type 3 hits allow to suppress atmospheric muon background at a level of signal from astrophysical neutrinos. Cumulative energy distributions of MC events from atmospheric muons after applying different cuts on number of type 2 pulses in event are shown in Fig. 5 left panel. Also shown is cumulative energy distribution of events expected from astrophysical flux with $E^{-2.46}$ spectrum and IceCube normalization. Expected cumulative energy distributions from astrophysical diffuse flux and atmospheric muons are equal at 40 TeV for $N_{type2} <1$ case and at 100 TeV for $N_{type2} <2$.

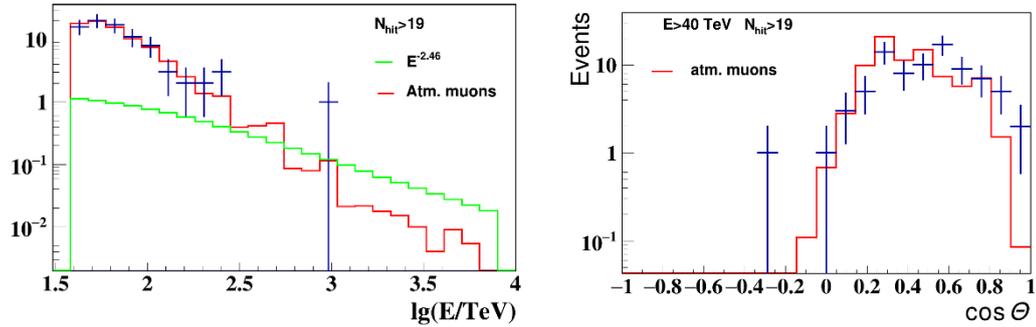

**Figure 3.** *Left*: Energy distribution of data events with $N_{hit} > 19$ and $E > 40$ TeV. Also shown are expected distributions of background events from atmospheric muons (red histograms) and from astrophysical neutrino flux with $E^{-2.46}$ spectrum (green histogram). *Right*: Cosine of zenith angle distribution of data events with $N_{hit} > 19$ and $E > 40$ TeV and expected distribution from atmospheric muons (red histograms).

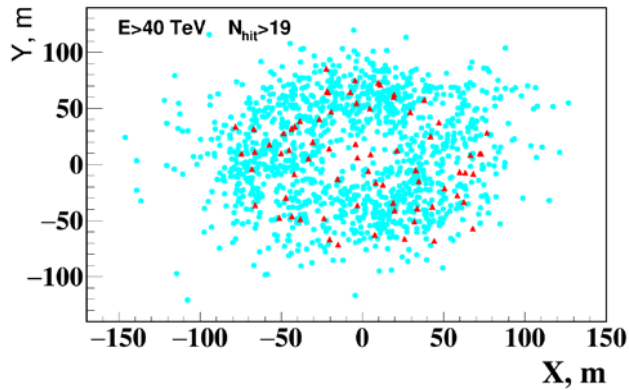

**Figure 4.** Reconstructed cascade vertices in horizontal plane for data (red triangles) and most bright cascades of simulated muon bundles (blue dots).

The following final cuts were applied to sample of 72 cascade-like events with $E > 40$ TeV and $N_{hit} >19$ for selection of events from astrophysical neutrinos: ($N_{type2}=0$ and $E_{rec} > 60$ TeV) or ($N_{type2}=1$ and $E_{rec} \geq 100$ TeV). Equal numbers of signal and background events are expected in experimental sample after applying final cuts. Seven events fulfill all selection requirements. There is no statistically significant excess of total number of events was observed over expectation from atmospheric muon background. Cumulative energy distributions of experimental events (red crosses), events expected from astrophysical flux with $E^{-2.46}$ spectrum and IceCube normalization (green histogram), events expected from atmospheric muons (brown histogram) and sum of expected astrophysical and atmospheric muon events (black histogram) are shown in Fig. 5 right panel. A higher statistics is required for observation of the astrophysical neutrino flux.

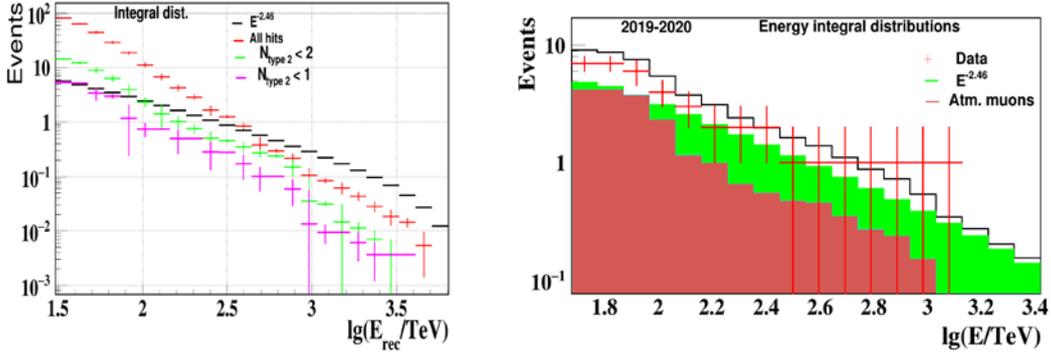

**Figure 5.** *Left:* Cumulative energy distributions of MC events from atmospheric muons after applying different cuts on number of type 2 pulses: red – without cuts, green – $N_{type2} < 2$, lilac - $N_{type2} < 1$. Also shown is cumulative energy distribution of events expected from astrophysical flux (black histogram). *Right*: Cumulative energy distributions of experimental events (red dots), events expected from astrophysical flux with $E^{-2.46}$ spectrum and IceCube normalization (green histogram), events expected from atmospheric muons (brown histogram) and sum of expected astrophysical and atmospheric muon events (black histogram).

Shown in Table 1 are parameters of the seven neutrino candidates: reconstructed energies, zenith and azimuthal angles as well as distances from cluster vertical axis and equatorial coordinates. Three, of seven astrophysical neutrino candidates are contained events. For contained cascades, in general, directional resolution (median value) is about 2.5° and for outside cascades about 4.5°. One event was reconstructed as a PeV scale cascade. Event GVD2019_1_114_N is reconstructed as upward going cascade with zenith angle θ = 109° and has very high probability to be neutrino induced cascade with signal-to-background ratio above 60%.

Table 1: Parameters of cascade events: reconstructed energy, zenith and azimuthal angles distances from cluster vertical axis, right ascension and declination.

|  | E,TeV | θ, degree | φ, degree | ρ, m | R.A. | Dec. |
|---|---|---|---|---|---|---|
| GVD2019_1_114_N | 91 | 109 | 92 | 49 | 45.1 | -16.7 |
| GVD2019_2_112_N | 1200 | 61 | 329 | 96 | 217.7 | 57.6 |
| GVD2019_2_153_N | 129 | 50 | 321 | 52 | 33.7 | 61.4 |
| GVD2019_3_663_N | 83 | 50 | 276 | 73 | 163.6 | 34.2 |
| GVD2020_3_175_N | 110 | 71 | 185 | 84 | 295.3 | -18.9 |
| GVD2020_3_332_N | 74 | 92 | 9 | 19 | 223.0 | 35.4 |
| GVD2020_6_399_N | 246 | 57 | 49 | 80 | 131.9 | 50.2 |

Recent works have shown that high-energy neutrinos of TeV to PeV energies are produced in active galactic nuclei. Specifically, within several parsecs from the supermassive black hole of radio-bright blazars with jets pointing towards us [9,10]. We base the following analysis on those conclusions, and present our preliminary results regarding the blazar-neutrino connection. For each Baikal-GVD high-energy neutrino event, we first list all VLBI-bright blazars in the complete flux density-limited sample [11] falling within the 50% uncertainty region. These regions have radii of 2.5 or 4.5 degrees, and typically contain several blazars.

We follow [9] and attempt to select the most likely neutrino sources by analyzing the variability information in the form of radio light curves. Rich light curves are provided by monitoring campaigns at RATAN-600 [12] and OVRO [13] telescopes, and are available for a large subset of the VLBI sample. We visually identify blazars that undergo major radio flares close in time to the cascade event as potential neutrino sources. Two events have one nearby object experiencing a major flare. Light curves of these blazars are shown in Fig. 6 and 7, together with sky maps surrounding corresponding GVD cascade events. The first is the J0301-1812 object. According to the multifrequency monitoring at RATAN-600, it was at the beginning of a major radio flare when a high-energy neutrino was detected from a direction 1.5 degree apart. The second example is J1938-1749, separated by 1.3 degrees from a cascade event. It was at the top of a flare, according to OVRO monitoring, and had the largest radio flux density observed over the whole monitoring duration.

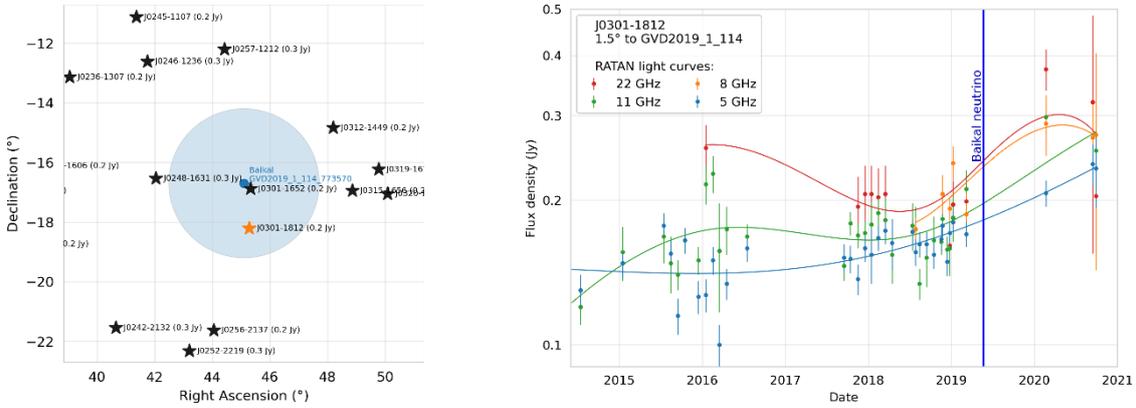

**Figure 6.** *Left*: Location of the GVD2019_1_114_N cascade event in the sky. The event itself is indicated by the blue dot in the center, with the light blue circle representing the 50% uncertainty region. VLBI-bright blazars from the complete sample are shown as stars with their names and radio flux densities to the side. The orange star corresponds to the blazar in the right panel. *Right*: Radio light curves from the RATAN-600 telescope for the J0301-1812 blazar, a potential neutrino source. This bright blazar undergoes a major flare, and the neutrino arrival coincides with the rise if this flare.

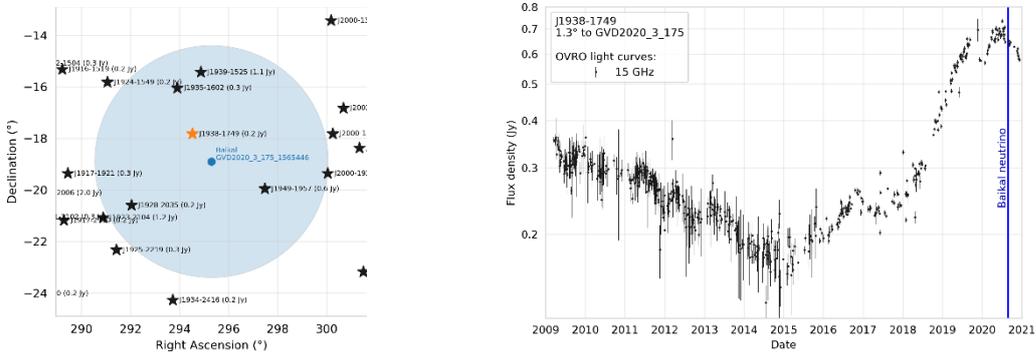

**Figure 7.** *Left*: Location of the GVD2020_3_175_N cascade event in the sky. See Fig. 6 caption for explanation of all symbols present. *Right*: The radio light curve from OVRO observations for the J1938-1749 blazar, a potential neutrino source. Neutrino arrived when this blazar was at its brightest over the whole observation period.

These results are preliminary, the proper statistical significance of spatial and temporal associations is yet to be determined. Nevertheless, we find them promising and encourage following up with analysis of events and sources presented here, as well as future neutrino detections at Baikal-GVD.

## 4. Conclusion

The ultimate goal of the Baikal-GVD project is the construction of a km$^3$-scale neutrino telescope with the implementation of about ten thousand photo-detectors. The first cluster in its baseline configuration was deployed in 2016. In total, eight clusters with 2304 OMs arranged at 64 strings are data taking since April 2021. Baikal-GVD in 2021 configuration is the largest neutrino telescope in the North at present. The modular structure of the Baikal-GVD design allows for studies of neutrinos of different origin at early stages of the construction. The analysis of data collected in 2019 and 2020 allows the selection of seven promising high-energy cascade events - candidates for events from astrophysical neutrinos. The commissioning of the first stage of the Baikal neutrino telescope GVD-1 with an effective volume 0.8 km$^3$ is envisaged for 2024-2025.

*We acknowledge the support by the Ministry of Science and Higher Education of Russian Federation under the contract FZZE-2020-0017. The work was supported by RFBR grant 128 20-02-00400. The CTU group acknowledges the support by European Regional Development Fund-Project No. CZ.02.1.01/0.0/0.0/16_019/0000766.*